\documentclass[preprint,prl,showpacs,amsmath,amssymb,superscriptaddress]{revtex4-1}

\usepackage{times}
\usepackage{graphicx}
\usepackage{dcolumn}
\usepackage{bm}
\usepackage{natbib}
\usepackage{color}
\usepackage{float}

\begin{document}

\preprint{APS/123-QED}

\title{Transient Exchange Interaction in a Helical Antiferromagnet}

\author{M.C. Langner}
\affiliation{Materials Science Division, Lawrence Berkeley National Laboratory, Berkeley, CA 94720, USA}
\author{S. Roy}
\affiliation{Advanced Light Source, Lawrence Berkeley National Laboratory, Berkeley CA 94720, USA}
\author{A. F. Kemper}
\affiliation{Computational Sciences Division, Lawrence Berkeley National Laboratory, Berkeley CA 94720, USA}
\author{Y.-D. Chuang}
\affiliation{Advanced Light Source, Lawrence Berkeley National Laboratory, Berkeley CA 94720, USA}
\author{S. K. Mishra}
\affiliation{Advanced Light Source, Lawrence Berkeley National Laboratory, Berkeley CA 94720, USA}
\author{R. B. Versteeg}
\affiliation{Advanced Light Source, Lawrence Berkeley National Laboratory, Berkeley CA 94720, USA}
\author{Y. Zhu}
\affiliation{Materials Science Division, Lawrence Berkeley National Laboratory, Berkeley, CA 94720, USA}
\author{M.P. Hertlein}
\affiliation{Advanced Light Source, Lawrence Berkeley National Laboratory, Berkeley CA 94720, USA}
\author{T.E. Glover}
\affiliation{Advanced Light Source, Lawrence Berkeley National Laboratory, Berkeley CA 94720, USA}
\author{K. Dumesnil}
\affiliation{Institut Jean Lamour (UMR CNRS 7198), Universit\'{e} de Lorraine, Vandoeuvre les Nancy, F-54500 France}
\author{R. W. Schoenlein}
\affiliation{Materials Science Division, Lawrence Berkeley National Laboratory, Berkeley, CA 94720, USA}

\date{\today}

\begin{abstract}
We have performed time-resolved resonant x-ray scattering studies in the Lanthanide metal Dy to reveal the dynamic response of the helical order exchange coupling to injection of unpolarized spins. The observed spin dynamics are significantly slower than that exhibited by the ferromagnetic phase in Lanthanide metals and are strongly dependent on temperature and excitation fluence. This unique behavior results from transient changes in the shape of the conduction electron Fermi surface and subsequent scattering events that transfer the excitation to the core spin.\end{abstract}

\maketitle

\chapter{}
Lanthanide metals exhibit a variety of magnetic phases due to competition between spin-orbit coupling, magneto-elastic effects, and long-range exchange coupling mediated by the indirect RKKY (Ruderman-Kittel-Kasuya-Yoshida) interaction  \cite{coqblin}.  The nature of the exchange interaction creates a composite spin system comprised of the closely coupled conduction and the core electron spins that account for the majority of the magnetic moment.  Helical or conically ordered phases, where the magnetic structure is characterized by a non-zero ordering wavevector, are created by competing symmetric and antisymmetric long-range exchange interactions, where the exchange interactions are determined by the spatial distribution of the conducting electron wave-functions.

Ultrafast demagnetization mechanisms have been previously explored in transition metal and rare earth magnets using all-optical and x-ray dichroism techniques \cite{ultrafastreview, REDurr, REWeinelt, REArpes, REEsch}. These studies have focused on the angular momentum transfer between core spins, conducting spins, and lattice, and addressed the dynamics of the uniform ferromagnetic phase.  Ultrafast optical pump/x-ray probe measurements reveal the  coupling mechanisms between the constituent spin systems in the Lanthanide magnets by observing the dynamics of the core spins in response to excitation of the conducting electrons responsible for the exchange interaction.  An as yet unexplored aspect of these magnetic systems are the dynamics in the helical phase, where transiently altering the conduction electron distribution can have a profound influence on the long-range magnetic structure.

In this study, we measure the dynamics of the inner shell f-electron spin helix in Dysprosium in response to transient injection of conduction electrons with unpolarized spins. We observe a reduction in the amplitude of the helical order parameter and a shift in the helical wavevector q on disparate time-scales that are strongly dependent on both the pump fluence and sample temperature. Notably, the dynamics in the helical antiferromagnetic phase (HAF) of Dy are significantly slower than those observed in the ferromagnetic (FM) phase of other Lanthanide metals \cite{REDurr, REWeinelt, REEsch}.  We attribute these anomalous dynamics to the relationship between the wave vector of the electronic excitation $k$ and the wave vector of the core magnetic ordering $q$.  In the FM phase, the electronic excitation at $k = 0$ is closely coupled to the magnetic ordering at $q = 0$.  In the HAF phase, the core spin helix is concomitant with a nesting of the Fermi surface (FS). Initial ultrafast scattering of spins from the $k=0$ excitation does not directly change the energy-minimizing configuration of the system, and instead the helical dynamics are driven by changes to the shape of the FS and subsequent changes to the electron distribution through scattering events. 

The spin helix was investigated in an epitaxially grown yttrium(Y)/dysprosium(Dy)/yttrium(Y) multilayer film that exhibits a second-order phase transition to a helical antiferromagnetic phase below $T_{N}$ = 180 K \cite{Karine}. A magnetostriction driven first order phase transition to a ferromagnetic phase occurs at $T_{C}$ = 60 K, with a critical temperature reduced from the bulk $T_{C}$ by the strain induced by the underlying yttrium layer. Between $T_{N}$ and $T_{C}$, the pitch of the spin helix, $\theta = q a$ (where a is the lattice constant), changes continuously with temperature, from 46 degrees (q=2.24 nm$^{-1}$) at $T_{N}$ to 30 degrees (q=1.46 nm$^{-1}$) at $T_{C}$ \cite{Sujoy}. 

The helical order parameter was was probed using resonant x-ray scattering, which provides a direct probe of the core-spin helical ordering through direct optical transitions between atomic core levels and the valence f-orbital \cite{Sujoy}. The sample is excited with an optical pump pulse with a photon energy of 1.5 eV and a duration of 100 fs.  The 20 nm top Y layer absorbs $\sim$ 90 \% of the pump energy, creating a hot electron distribution in the Y layer and resulting in ultrafast injection of unpolarized spins into the 500 nm Dy film via nonequilibrium diffusion \cite{REDurr, REWeinelt, SuperDiffusive}.  Time-resolved diffraction measurements were performed at beamline 6.0.2 at the Advanced Light Source, Lawrence Berkeley National Laboratory, utilizing a probe energy of 1290 eV, resonant with the Dy M5-edge, with a probe pulse duration of 70 ps.

\begin{figure}[H]
\includegraphics[width=3.2in]{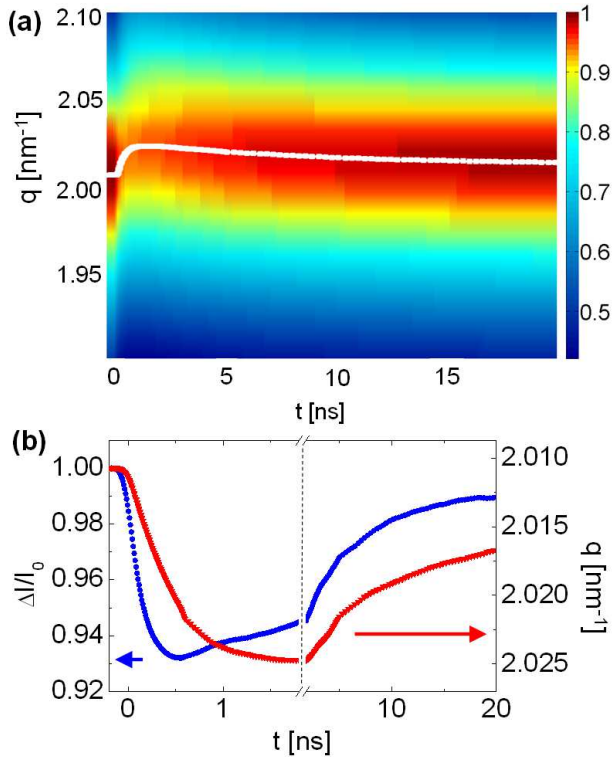}
\caption{(a) Colormap of the diffraction peak intensity as a function of time delay.  The white line indicates the time-dependence of the scattering wave-vector.  (b) Scattering peak intensity and scattering wave vector as a function of time delay for the data in panel (a).}
\label{fig:fig1}
\end{figure}

Figure \ref{fig:fig1} shows the spiral diffraction peak with a pump fluence of 0.66 mJ/cm$^2$ at 105 K as a function of pump-probe time delay.  The dynamics of the diffraction peak, consisting of the onset of the excitation and subsequent recovery, are faster than the corresponding shifts in scattering wavevector.  The initial response of the helix is characterized by a reduction of diffraction intensity (I) occurring on a 200 ps time scale, and an increase of the peak wave-vector (q) occurring on a 1 ns time scale. Both q and I recover on time scales of several ns.  The parameters I and q are determined by fitting the diffraction peak with a Lorenztian function. There is no observable change in the peak width, which is likely limited by the penetration depth of the x-ray probe, and the dynamics of the diffraction signal can therefore be completely characterized by I(t) and q(t).  

Figure 2 shows the time-dependence of I and q as a function of pump fluence at 105 K.  The excitation and recovery dynamics become slower with increased fluence for both the parameters. The maximum reduction in the diffraction intensity ($\Delta I/I_{0}$) is linear with fluence, while the maximum shift in wavevector shows a saturation-like behavior at early time delays with fluences above 1.5 mJ/cm$^2$.

\begin{figure}[H]
\includegraphics[width=2.8in]{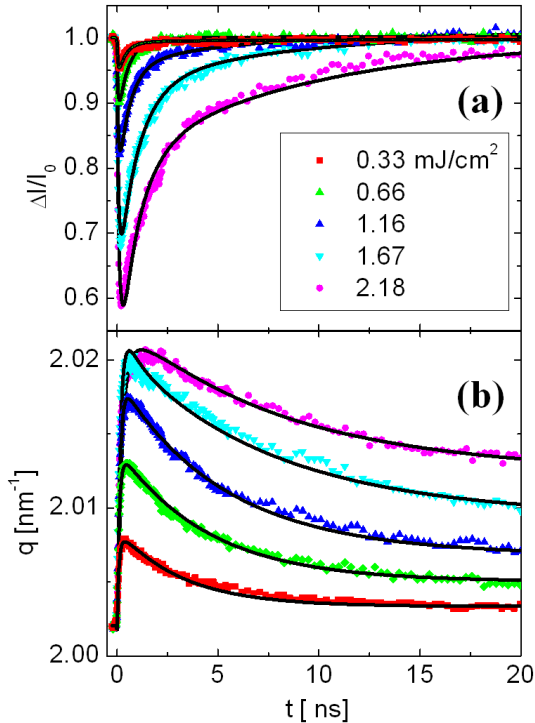}
\caption{Time-dependence of the loss in diffraction peak intensity (a) and change in q-vector (b) as a function of pump fluence.  Solid lines indicate fits to the model described in the text.}
\label{fig:fig2}
\end{figure}

At long time scales ($>$ 20 ns), the shift in q is linear in fluence and consistent with an increase in temperature of the sample. On this time scale both I and q can be parameterized by an increase in the temperature, consistent with measured static values. At intermediate temperatures, static measurements show little variation in the diffraction efficiency, with a monotonic decrease in q as the sample is cooled from $T_{N}$ to $T_{C}$ \cite{Sujoy}. The large transient photo-induced reduction in I, coupled with only a moderate increase in q, indicates that the dynamics observed on time-scales faster than $\sim$ 20 ns are non-thermal, and the helical system at short time scales cannot be described with an effective spin temperature.

The microscopic origin of the HAF dynamics is clarified by considering the relationship between the core spin ordering and the conduction electron FS.  In equilibrium, the FS has extended regions in which electrons can scatter from $\phi \left( k \right)$ to $\phi \left( k + q \right)$, which combined with the exchange interaction between core and conduction spins, leads to helical ordering of the core spins with wave vector q \cite{Keeton}.  The temperature dependence of q results primarily from the dependence of the helical ordering energy gap on thermal fluctuations of the basal plane magnetization, leading to a re-shaping of the FS as the thermal fluctuations are frozen out \cite{EW1, EW2, YW, ARPES}.

Magneto-optical Kerr measurements indicate a sub-ps injection of hot unpolarized electrons from the Y layer \cite{Supplemental}, consistent with dynamics observed for the same process in other magnetic materials \cite{REDurr, REWeinelt, SuperDiffusive}.  The hot electrons create an excitation at $k=0$ in the conduction spins, perturbing the magnetic ordering but preserving the FS nesting q.  The induced disorder of the conduction spins reduces the number of quasi-elastic $k \rightarrow k + q$ scattering events where spin angular momentum is conserved, and increases $k \rightarrow k + q$ scattering involving changes in angular momentum.  The core/conduction exchange couples this angular momentum scattering to the core helix, propagating spin disorder to the core spins, but not initially changing the favored helical wave-vector.
  
\begin{figure}[H]
\includegraphics[width=3.0in]{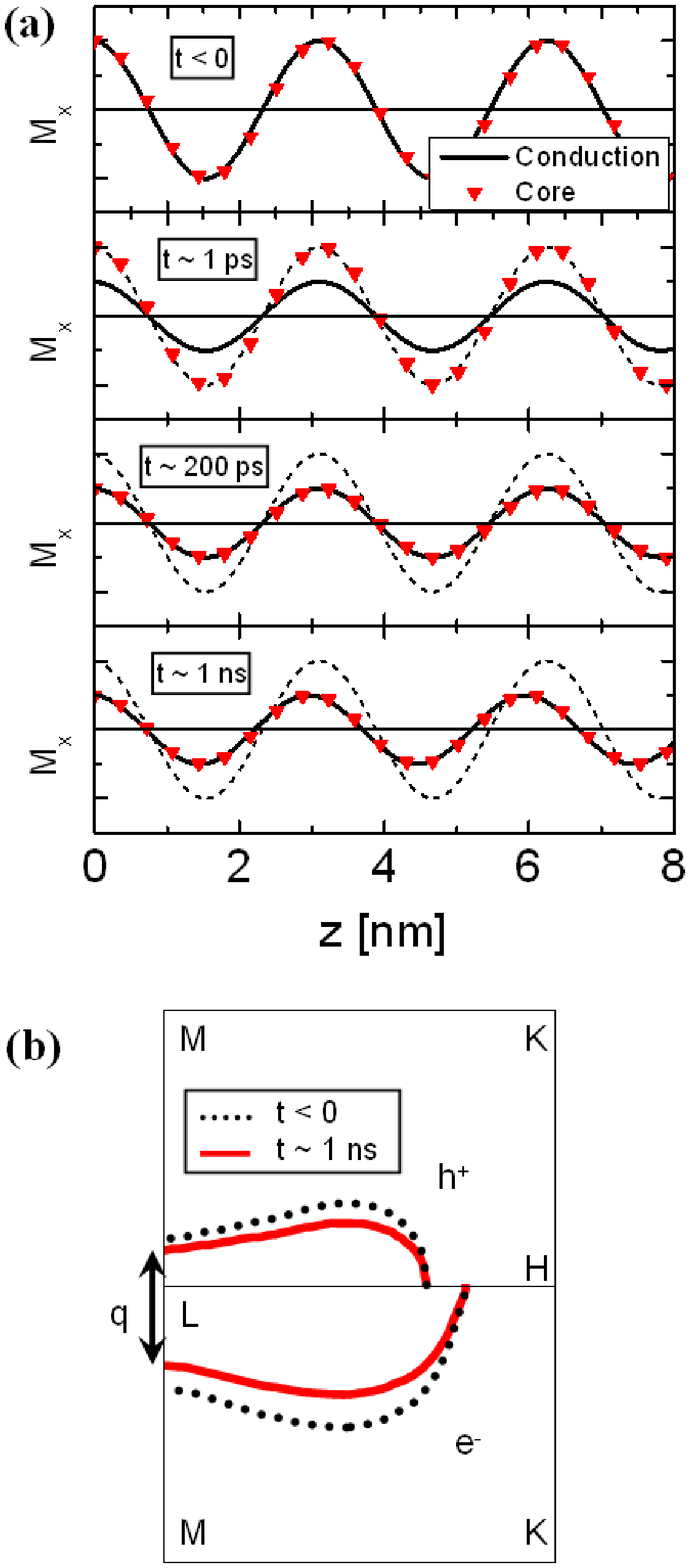}
\caption{(a) Diagram of real-space spin ordering of core and conduction electrons for a spiral with $q \| z$.  The y-axis units are arbitrary.  (t $<$ 0) Equilibrium distribution (t $\sim$ 1 ps) Photoexcitation of conduction electrons, followed by (t $\sim$ 200 ps) excitation of core spins, and (t $\sim$ 1 ns) subsequent shift in q.  The dotted lines show the equilibrium distribution for reference. (b) Corresponding FS diagram.  Adapted from \cite{Keeton}}
\label{fig:fig3}
\end{figure}

The response of the core-spins to the perturbed conduction electron distribution mimics the reduction in the magnetization due to thermal excitations, and the equilibrium wave vector q for the FS nesting and core helix changes due to the dependence of q on the basal plane magnetization.  This process is diagrammed in figure \ref{fig:fig3}.  The shift in the nesting vector of the FS creates a mismatch between the FS and the electron distribution.  The electron distribution relaxes via inelastic scattering events that couple electrons with excess energy (above E$_{Fermi}$) with hole states below E$_{Fermi}$, with transfer of spin angular momentum occurring through exchange and spin-orbit coupling interactions.  This situation is analogous to the breathing Fermi surface model used to describe damping of the precession in ferromagnets, in which the dynamical pointing of the ferromagnetism alters the FS through spin-orbit coupling \cite{BFS1, BFSNonCol, BFSNonCol2}. The excitation time-scales observed in the helical system are similar to damping time-scales in other magnetic systems, limited by the scattering rate and efficiency of angular momentum transfer during scattering events.

The three temperature model is often invoked to describe the dynamics of magnetic systems in terms of energy transferred between electronic, lattice, and spin degrees of freedom. In a ferromagnet, a single temperature defines the state of a global, uniform magnetization, whereas in the HAF phase of Dy, the magnetic state is defined by both the strength of the order parameter and the helical wavevector. Dynamically, the relation between I and q differs from the I/q relationship in equilibrium, thus simple models based on an effective spin temperature are inadequate for describing this system.

Instead, we employ a Gross-Pitaevski (GP) model, wherein the dynamics of I and q are coupled, resulting from changes in the nested FS and subsequent interactions between core and conducting electrons \cite{Li, GP}. We use a Hamiltonian given by

\begin{equation}
\label{eq:Heff}
H = \frac{J_{1}}{2a} \int{d^{3} x} \left[- \frac{\theta{}}{2} \left( \nabla m \right)^{2} + \frac{a}{4} \left( \nabla^{2} m \right)^{2} \right],
\end{equation}

where $m$ is the strength of the order parameter of the HAF, proportional to the experimentally measured I. This Hamiltonian contains the exchange terms $J_{1}$ and $J_{2}$, which are the effective nearest and next-nearest neighbor coupling between core spins that stabilize the helical phase.  $J_{2}$, appears through $\theta$, which is the equilibrium turn angle of the helix given by $cos(\theta) = J_{1}/4J_{2}$.  The helix is the lowest energy state of this system when $J_{1}$ and $J_{2}$ have opposite sign.

For a one-dimensional helical magnetic structure with wave-vector $q$, the above Hamiltonian leads to an effective free-energy given by:

\begin{equation}
F = - m^2 \alpha \left( T \right) + \beta m^4 + \frac{1}{2} J_{1} a^2 m^2 \left( - \frac{\theta^2 q^2}{2} + \frac{a^2 q^4}{4}\right)
\label{eq:FE}
\end{equation}.

The first two terms, with factors $\alpha$ and $\beta$, are the lowest-order terms in the free energy expansion that stabilize the order parameter for $T < T_{N}$, and $a$ is the lattice constant. Changes to the free energy of the core helix arising from excitation of the conduction electrons are considered through the parameters $J_{1}$  and $J_{2}$ .  Note that $\theta$ and $q$ are retained as variables to distinguish the equilibrium helix turn angle from the measured dynamic variable q.

The dynamics can be calculated by using the Hamiltonian to describe an effective action of the helical system parameterized by $m$ and $q$.  From equation \ref{eq:FE}, we can write the equations of motion as

\begin{align}
\gamma_{m} \dot{m} &= 2 \alpha m - 4 \beta m^3  - J_{1} m a^2 \left( - \frac{q^2 \theta^2}{2} + \frac{a^2 q^4}{4} \right) \label{eq:motion1} \\
\gamma_{q} \dot{q} &= -\frac{1}{2} J_{1} m^2 a^2 q (- \theta^2 + a^2 q^2 ) \label{eq:motion2}.
\end{align}

Due to the absence of oscillations in the data, we neglect the second derivative terms in the above equation.  Additionally, we have introduced phenomenological damping terms with parameters $\gamma_{m}$ and $\gamma_{q}$ to account for relaxation of the system back to equilibrium. 

The behavior of $m$ and $q$ with different rheonomic constraints provides insight into the origins of the dynamics of the spin helix. Within the GP model, the dynamics of the parameters $J_{1,2}$ emulate the dynamics of the FS, and the damping parameters model the scattering mechanisms that drive the spin-ordering to match the FS.  The observed dynamics I and q are described by introducing time-dependent parameters $J_{1}(t)$ and $J_{2}(t)$ into equations \ref{eq:motion1} and \ref{eq:motion2} such that the derived $m$ and $q$ match the data.  

We choose $J_{1}(t)$ and $J_{2}(t)$ to be consistent with our microscopic picture in the following ways.  The uniform (k = 0) excitation of the conduction electrons is modeled as a proportional reduction in both exchange parameters, such that initially the equilibrium q-vector is unchanged.  We introduce two recovery time-constants to describe the observed recovery times in the data.  We introduce parameters $\sigma$ and $\rho$ to account for the symmetric and asymmetric recovery amplitudes of the exchange constants.  Within this model, the exchange parameters vary according to 

\begin{align}
\frac{\delta J_{1}(t)}{J_{1}(0)} &= \delta J_{init}((1 - \sigma - \delta J_{T,1}) e^{-t/\tau_{a}} + \sigma e^{-t/\tau_{b}} + \delta J_{T,1}) \\
\frac{\delta J_{2}(t)}{J_{2}(0)} &= \delta J_{init}((1 - \sigma - \rho_{2}) e^{-t/\tau_{a}} +  \sigma e^{-t/\tau_{b}} + \rho_{2} e^{-t/\tau_{2}}).
\end{align}

The onset of the reduction in $J_{1,2}$ is instantaneous to be consistent with observed optical data [13], and the slow onset of the reduction in $m$ is limited by the damping parameter $\gamma_{m}$. Shifts in $q$ are treated as partial asymmetric recoveries of either exchange parameter.  The black lines in figure 2 show fits to the data using $m(t)$ and $q(t)$ calculated with the GP model.

\begin{figure}[H]
\includegraphics[width=3.0in]{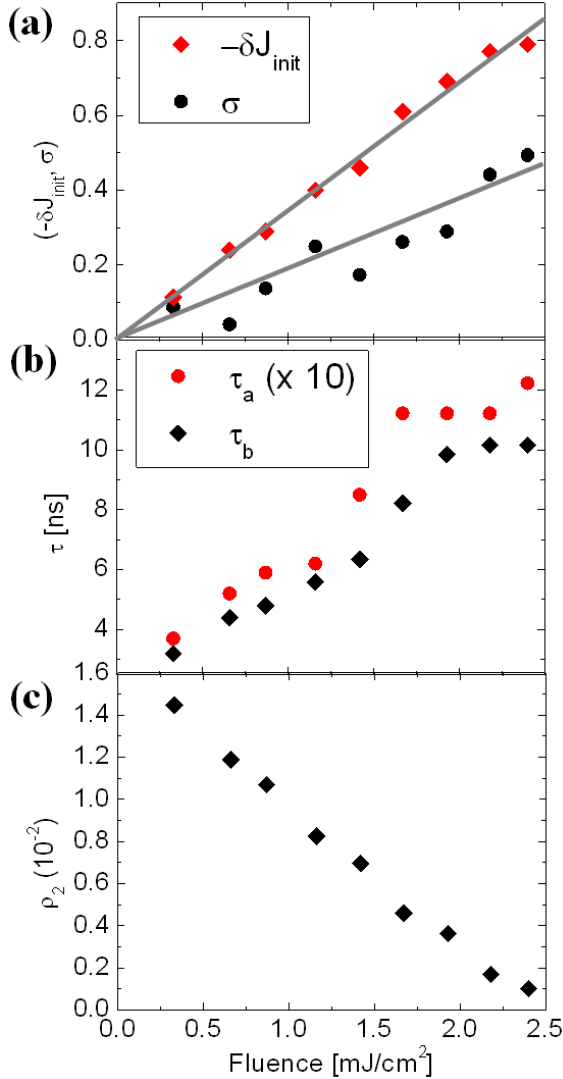}
\caption{Magnitudes and time-scales of changes in exchange parameters $J_{1}$ and $J_{2}$ used to describe the fluence dependence of the data.  (a) Amplitudes of initial reduction and proportional recovery of the exchange parameters.  Note that $\delta J_{init}$ is negative, and the line indicates a linear reduction in the exchange parameters with fluence.  (b) Proportional recovery timescales.  (c) Non-proportional recovery in $J_{2}$ leading to the shift in $q$.  $\rho_{2}$ becomes smaller with fluence, leading to a saturation-like behavior for the shift in spiral wavevector.}
\label{fig:fig4}
\end{figure}

Figure \ref{fig:fig4} (a) shows the initial reduction in the exchange parameters $\delta J_{init}$ and the recovery amplitude $\sigma$.  The initial reduction in the exchange is linear with fluence, as would be expected from an effective exchange proportional to the spin ordering and an injected unpolarized spin population that scales with the fluence.  The initial change in the exchange constants are equal for $J_{1}$ and $J_{2}$ and therefore do not lead to a shift in the spiral wavevector.  The time-scales for the recovery, $\tau_{a}$ and $\tau_{b}$ are shown in figure \ref{fig:fig4} (b).

The initial shift in $q$ occurs through a relatively fast recovery of $J_{2}$ through the term $\rho_{2}$.  This parameter is shown in figure \ref{fig:fig4} (c).  The shift in q for time delays greater than 20 ns is treated through $J_{T,1}$, which is a small remnant reduction in $J_{1}$ representing an increased sample temperature.  Note that $\{ J_{T,1}, \rho_{2} \} \ll \{ \delta J{i}, \sigma \}$; the terms representing anti-symmetric dynamics in $J_{1}$ and $J_{2}$ that shift the spiral $q$ are much smaller than the symmetric changes in the exchange constants, resulting in overall dynamics of $J_{1}$ and $J_{2}$ that are nearly symmetric \cite{Supplemental}.

\begin{figure}[H]
\includegraphics[width=3.0in]{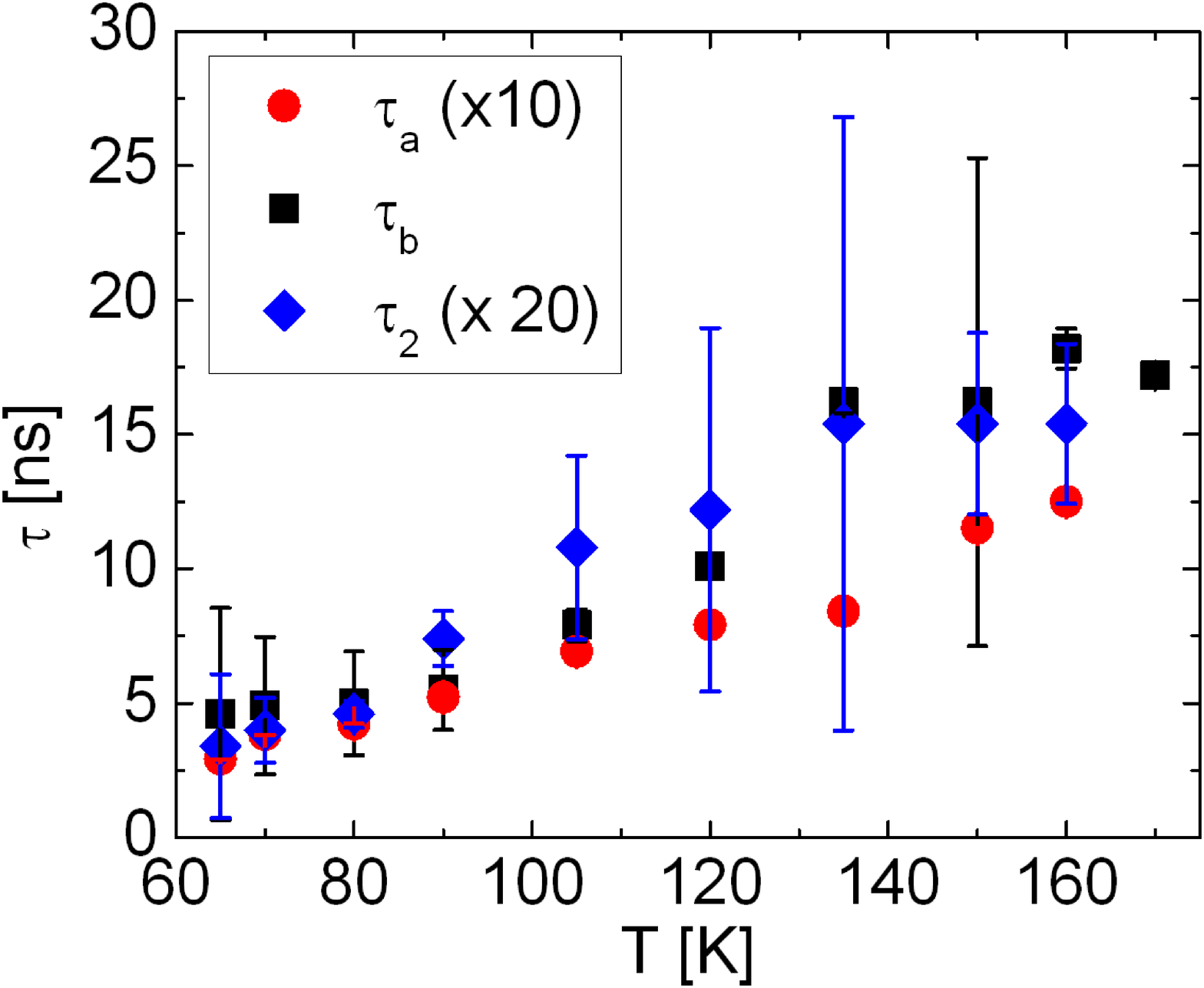}
\caption{Time-scales of exchange parameters $J_{1}$ and $J_{2}$ as a function of temperature.}
\label{fig:fig5}
\end{figure}

The modeled time-constants become slower with increasing fluence (figure \ref{fig:fig4}(b)) and also become slower as T is increased from $T_{C}$ to $T_{N}$ (figure \ref{fig:fig5}). Both of these trends suggest a link between the magnetic ordering and the recovery time-scales of the spin helix.  This behavior is consistent with changes in the shape of the FS, with a corresponding reduction in the effective exchange coupling between core spins, resulting from a reduction of the basal plane magnetization $m$ \cite{EW1, EW2, YW, ARPES}.

The equilibrium helical phase results from the interaction between the core spins and the nested FS of the conduction electrons; the phenomenological GP model provides a short-hand to account for dynamics of the nested FS through $J_{1,2}$ and subsequent electron population scattering events through $\gamma_{m,q}$. By considering changes to the shape of the free energy surface, the GP model provides an accurate description of the helical motion with fluence and temperature dependence that is qualitatively consistent with both the static and optical dynamic measurements. We observe a linear reduction in the exchange coupling with laser fluence, indicating a direct excitation of the conduction electrons, and fluence and temperature dependence of the recovery time scales consistent with an effective exchange coupling that scales with $m$.

In summary, the dynamics of the helical phase in response to transient unpolarized spin injection differ significantly from those in the ferromagnetic phase due to the relationship between the core spins and conduction electron FS nesting.  FM phase dynamics result from close coupling of the $k = 0$ excitation to the core spins through spin-orbit coupling and short range exchange interactions.  The dynamics of the helical phase result from indirect excitation to the finite wave-vector ordering through a fundamentally different process, analogous to a damping mechanism, that transfers angular momentum between the excited conduction electrons and core spins. 

\begin{acknowledgments}
The work at LBNL, including experiments at ALS, was supported by the Director, Office of Science, Office of Basic Energy Sciences, of the U.S. Department of Energy under Contract No. DEAC02-05CH11231.
\end{acknowledgments}

\pagebreak

\chapter{}
\setcounter{figure}{0}
\makeatletter 
\renewcommand{\thefigure}{S\arabic{figure}}

\begin{center}
Supplementary Material for\\ ``Transient Exchange Interaction in a Helical Antiferromagnet''
\end{center}

\section{Optical Pump-Probe Measurements}

\begin{figure*}
\includegraphics[width=5in]{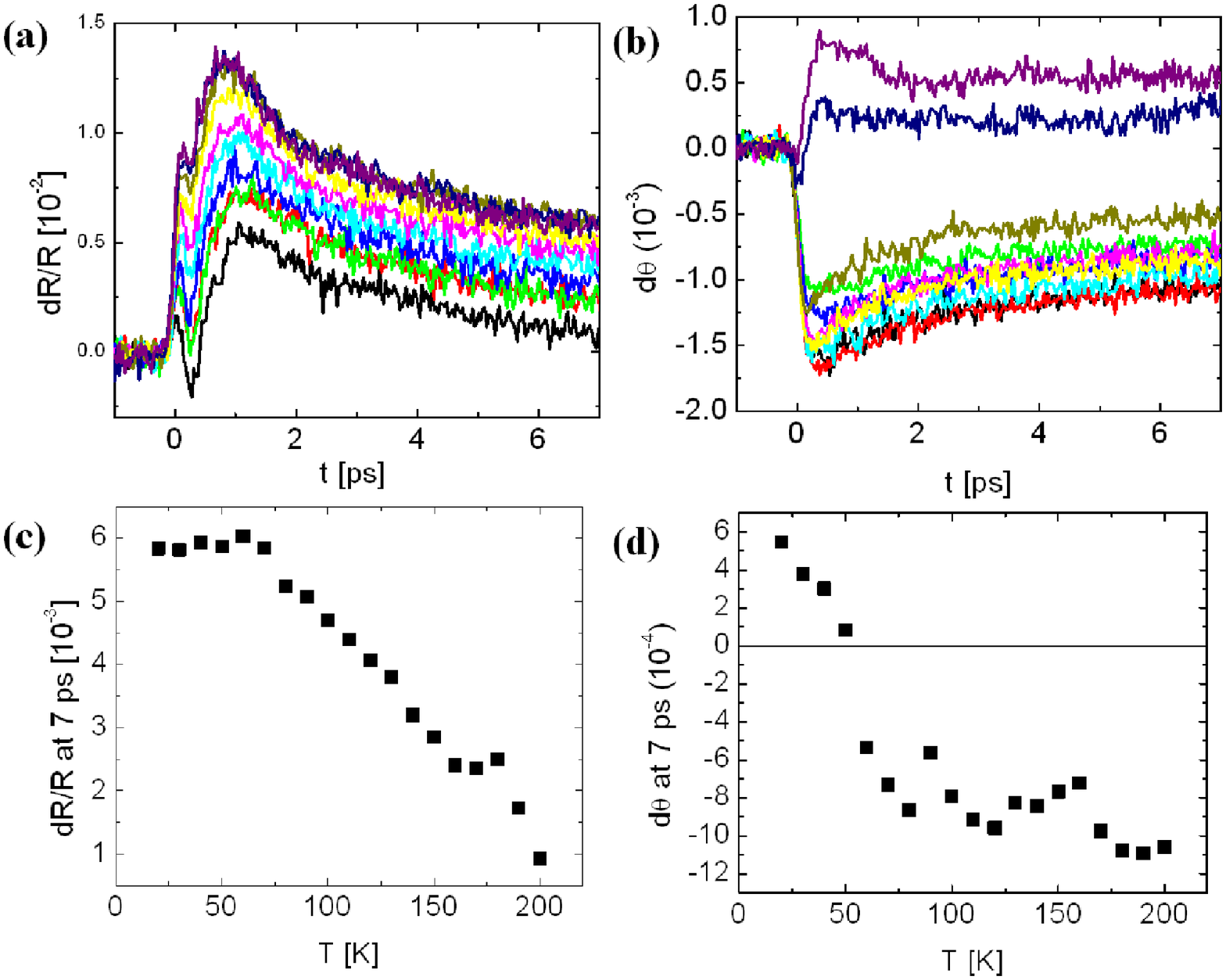}
\caption{Optical pump-probe data of the Y/Dy/Y thin film.  (a) dR/R as a function of temperature from 20 K (top, purple) to 200 K (bottom, black).  (b) ($\theta_{K}$) as a function of temperature from 20 K (top, purple) to 200 K (bottom, black).  (c,d) Temperature dependence of the amplitude of dR/R (c) and ($\theta_{K}$) (d).}
\label{fig:S1}
\end{figure*}

The photo-induced change in reflectivity (dR/R) and magnetooptic Kerr (MOKE) angle ($\theta_{K}$) for the Y/Dy/Y stack are shown in figure \ref{fig:S1}.  The reflectivity and MOKE measurements were recorded with a probe energy of 3.0 eV with an excitation energy of 1.5 eV.  The MOKE signal shows a clear change in amplitude at the ferromagnetic transition temperature on times scales of $<$ 1 ps, indicating that the 1.5 eV optical pulse excites the conduction-level magnetism on time-scales similar to those observed in previous ultrafast measurements of rare-earth magnetism \cite{ultrafastreview, REDurr, REWeinelt, REArpes, REEsch}.  A sharp change in amplitude at $T_{C}$ is not observed in the reflectivity measurements.

\section{Lattice Strain}

The c-axis lattice constant changes with temperature, corresponding to the changes in the spiral ordering \cite{Karine}.  The lattice constant is reduced by $\delta c/c = 0.003$ as the sample is heated through the helical phase, making the observed shifts in the magnetic wavevector q with laser excitation too large to be attributed to a thermal change in the lattice constant.  In our analysis of the dynamical data, changes in the lattice constant are considered indirectly through the effective exchange constants.

\begin{figure*}
\includegraphics[width=5in]{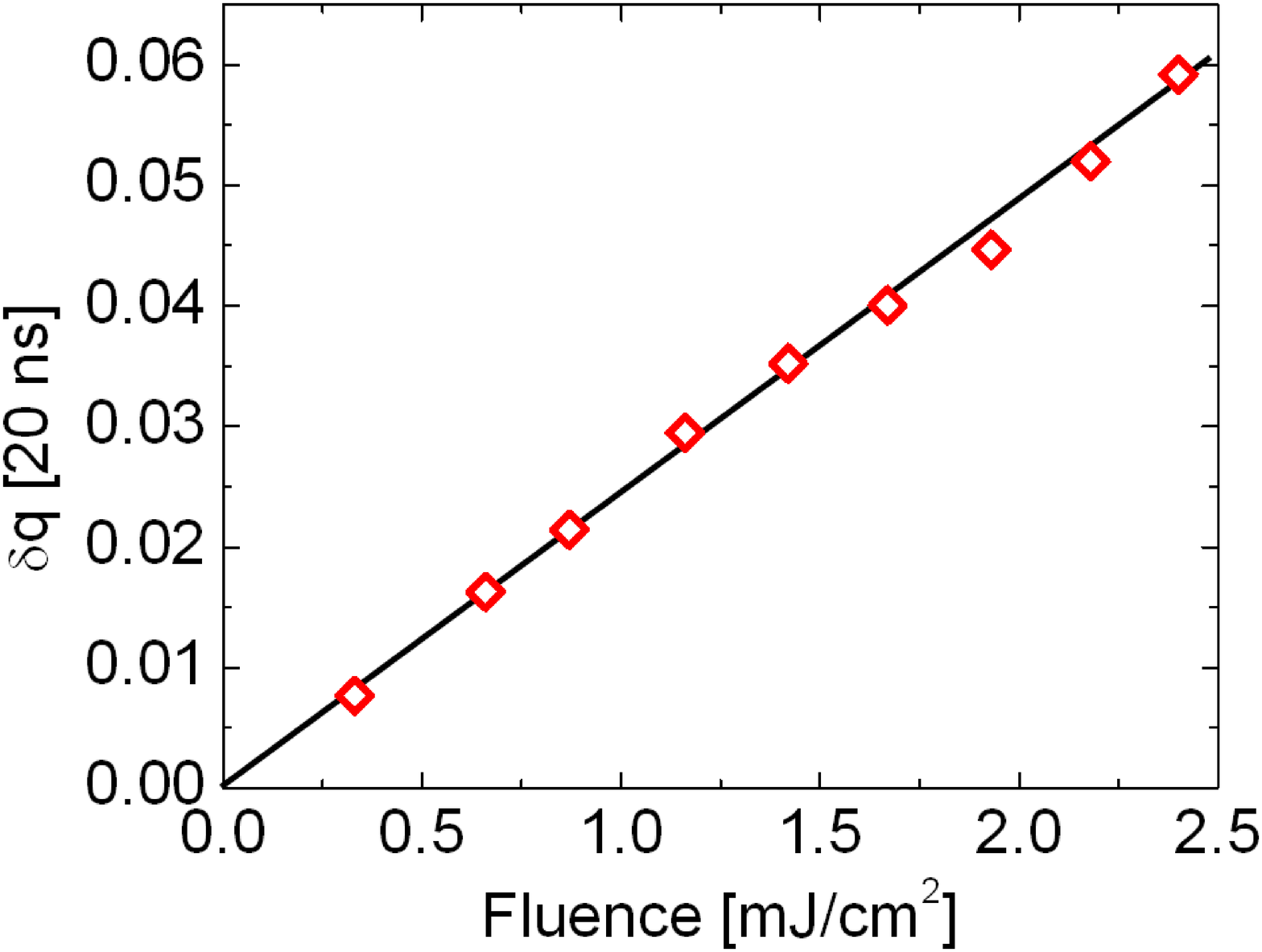}
\caption{Fluence dependence of shift in q-vector at 105 K.}
\label{fig:S1b}
\end{figure*}

On long time-scales ($>$ 20 ns ), the changes in the diffraction peak are consistent with an increase in the sample temperature, with little change in the peak intensity at intermediate temperatures and an increase in wavevector.  The increase in q-vector with fluence is shown in figure \ref{fig:S1b} at 105 K.  The linear dependence on fluence is expected for a regime in which the specific heat is roughly constant with photo-induced change in temperature with a temperature increase of $\sim$ 5 K at the highest fluence.  At times shorter than $ \sim $ 20 ns, the relatively large decrease in scattering intensity indicates that the dynamics are non-thermal.  

\section{Fits of Temperature-Dependent Data to Gross-Pitaevski Dynamics}

\begin{figure*}
\includegraphics[width=5in]{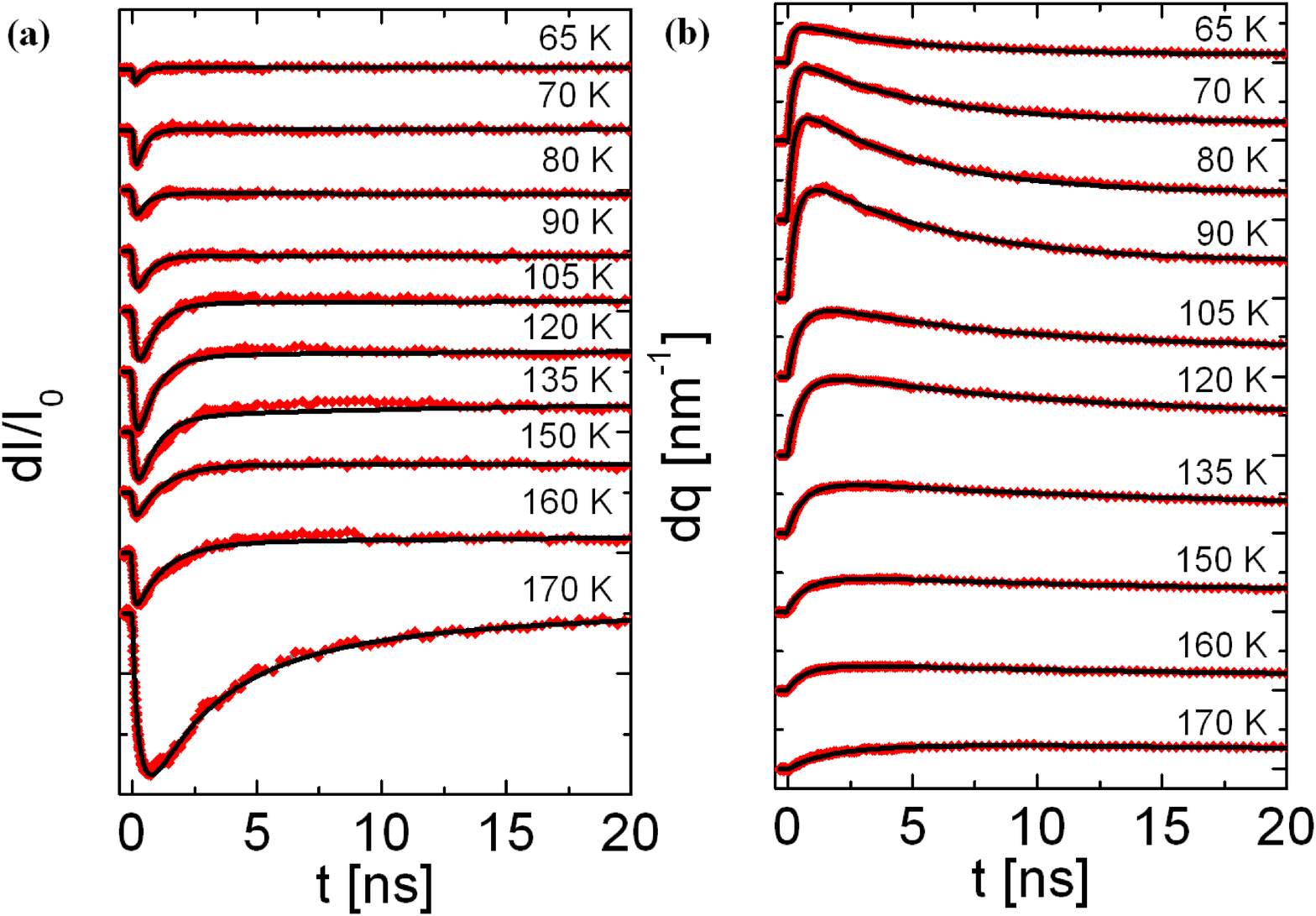}
\caption{Temperature dependence of (a) dI/I and (b) wavevector ($q$) for a pump fluence of 0.66 mJ/cm$^2$.  Lines are fits from the GP model described in the main text.}
\label{fig:S2}
\end{figure*}

\begin{figure*}
\includegraphics[width=3.4in]{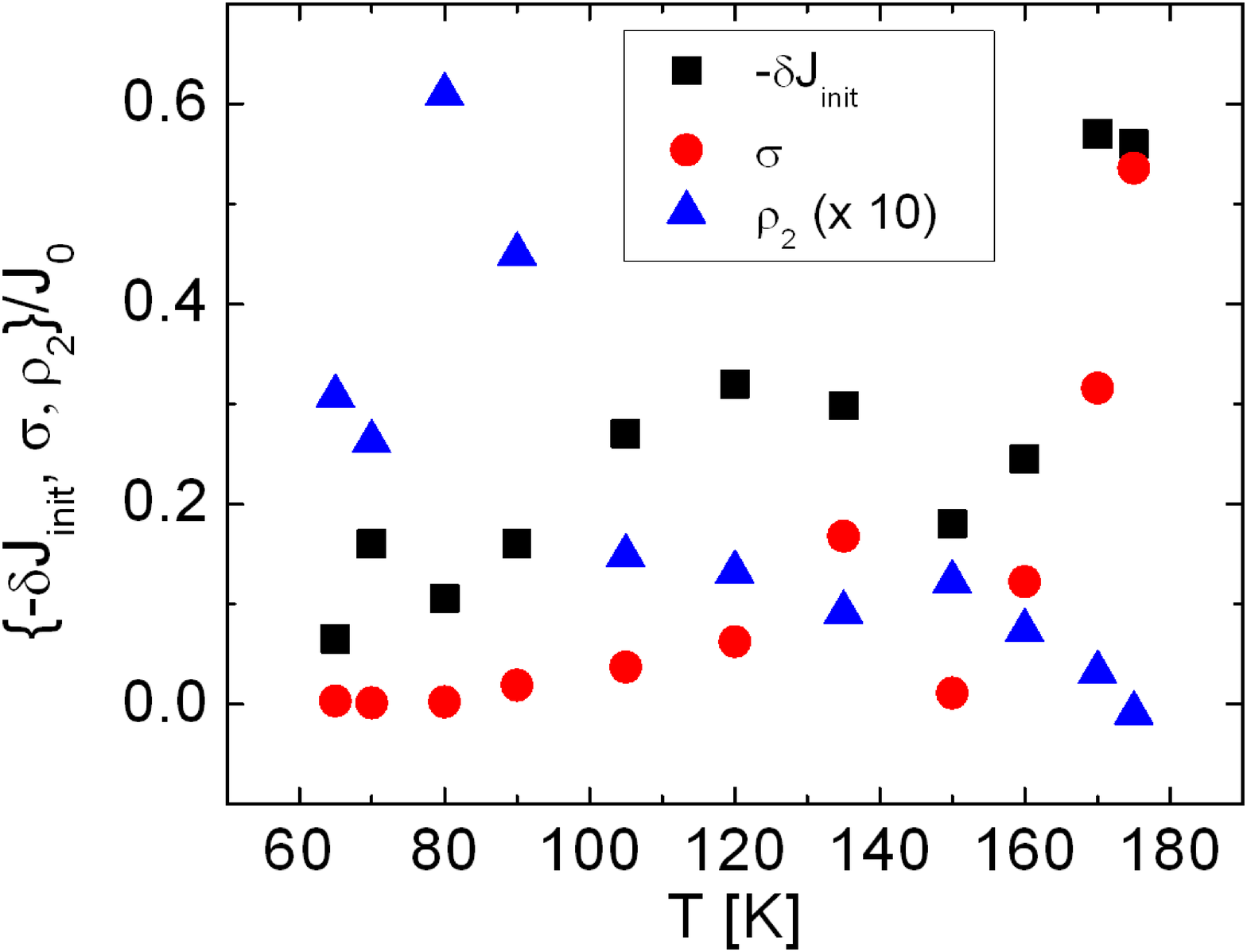}
\caption{Reduction amplitudes of $J_{1}$ and $J_{2}$ as a function of temperature.  The component amplitudes are described in the main text.}
\label{fig:S3}
\end{figure*}

Figure \ref{fig:S2} shows the temperature dependence of the spin-ordering peak dynamics.  The dynamics of the exchange constants become progressively slower at higher temperature (figure 5 of the main text), with little change in the reduction amplitude except near $T_{N}$, as shown in figure \ref{fig:S3}.

\section{Time-dependence of $J_{1}$ and $J_{2}$}

\begin{figure*}
\includegraphics[width=6in]{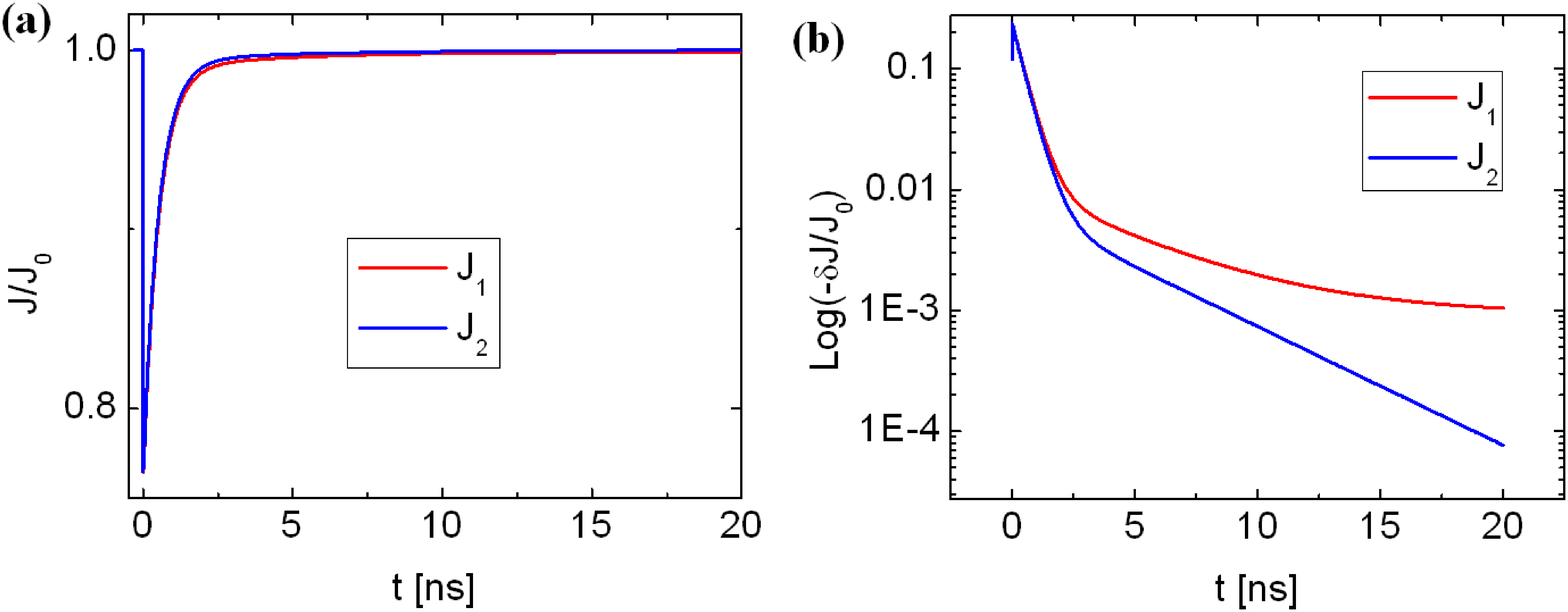}
\caption{(a) Normalized dynamics of $J(t)_{1}$ and $J(t)_{2}$.  (b) Log plot of the normalized dynamics}
\label{fig:S4}
\end{figure*}

Figure \ref{fig:S4} shows the time dependence of $J_{1}$ and $J_{2}$ for data at 105 K with a pump-fluence of 0.66 mJ/cm$^2$.  The $J_{2}$ exchange parameter partially recovers on a relatively fast time-scale, initiating the shift in $q$, while a partially unrecovered $J_{1}$ is used to describe the change in $q$ due to an increase in temperature of the sample.

\end{document}